\documentclass{emulateapj}

\def\lax {\ifmmode{_<\atop^{\sim}}\else{${_<\atop^{\sim}}$}\fi}  
\def\gax {\ifmmode{_>\atop^{\sim}}\else{${_>\atop^{\sim}}$}\fi}  
\def\gtorder{\mathrel{\raise.3ex\hbox{$>$}\mkern-14mu
             \lower0.6ex\hbox{$\sim$}}}

\def\pl{{\sc pl}}

\def\chiq{$\chi^2$}

\def\nh{{$N_{\rm H}$}}

\def\I{{\em INTEGRAL}}
\def\S{{\em Swift}}

\def\0840{IGR~J08408--4503}
\def\be{\begin{equation}}
\def\ee{\end{equation}}

\begin{document}

\title{IGR J08408--4503: a new recurrent Supergiant Fast X-ray Transient}

\author{D. G\"{o}tz\altaffilmark{1}}\email{diego.gotz@cea.fr}
 \author{M. Falanga\altaffilmark{1,2}}
  \author{F. Senziani\altaffilmark{3,4,5}}
\author{A. De Luca\altaffilmark{3}}
  \author{S. Schanne\altaffilmark{1}}
\author{A. von Kienlin\altaffilmark{6}}

\altaffiltext{1}{CEA Saclay, DSM/DAPNIA/Service d'Astrophysique,
  F-91191, Gif sur Yvette, France} 
\altaffiltext{2}{Unit\'e mixte de recherche Astroparticule et
  Cosmologie, 11 place Berthelot, 75005 Paris, France}  
\altaffiltext{3}{INAF--IASF Milano, via Bassini 15, I-20133 Milano, Italy}
\altaffiltext{4}{Universit\'e Paul Sabatier, 31062 Toulouse, France}
\altaffiltext{5}{Universit\`a di Pavia, Dipartimento di Fisica
Nucleare e Teorica and INFN--Pavia, via Bassi 6, I-27100 Pavia, Italy}
\altaffiltext{6}{Max-Planck-Institut f\"{u}r extraterrestrische Physik, Giessenbachstrasse,
D-85748 Garching, Germany}

\begin{abstract}
The supergiant fast X-ray transient \0840 was discovered by \I\ on May
15, 2006, during a bright flare. 
The source shows sporadic recurrent short bright flares, reaching a peak luminosity of
10$^{36}$ erg s$^{-1}$ within less than one hour. 
The companion star is HD 74194, an Ob5Ib(f) supergiant star
located at 3 kpc in the Vela region. We report 
the light curves and broad-band spectra (0.1--200 keV) of all the three flares of \0840\ detected up to now based on \I\ and \S\ data.
The flare spectra are well described by a power-law model with a high energy cut-off at $\sim$ 15 keV. The absorption column density during the flares was found to be $\sim 10^{21}$ cm$^{-2}$, 
indicating a very low matter density around the compact object.
Using the supergiant donor star parameters, the wind accretion conditions imply an orbital period of the order of one year, a spin period of the order of hours and a magnetic field of the order of 10$^{13}$ G.
\end{abstract}


\section{Introduction}
\label{sec:intro} 

Supergiant Fast X-ray Transients (SFXTs), consisting of a
wind accreting compact object and an OB supergiant donor star, spend
most of the time in a quiescent state, with X-ray 
luminosities of the order of $10^{32}-10^{33}$ erg s$^{-1}$. Sporadically
they go into outburst reaching luminosities of $10^{36}-10^{37}$ erg s$^{-1}$
\citep{sguera,zand,negueruela,smith}. The outbursts or
strong flaring activities are characterized by very short time
scales, i.e. from minutes to hours. 
SFXTs have been discovered only recently as a new class of X-ray sources, and may unresolved
fundamental questions remain open. The lack of confirmed orbital or
pulse periods complicates the determination of the system parameters as well as of 
the compact object nature. Up to now IGR~J11215-5952 alone shows regularly recurrent
outbursts every $\sim 330$ days, most likely linked to its orbital
period \citep{sidoli}. To account for the accretion process that causes the
short outbursts, it was proposed that the clumpy wind in
early-type stars could be captured by the compact
object producing the X-ray flares on the observed time scale
\citep{zand}. However, this scenario has yet to be well understood and confirmed for OB
stars in binary systems.

The number of known SFXTs has grown recently thanks to the International Gamma-Ray
Astrophysics Laboratory (\I), see e.g. IGR~J16465-4507,
IGR~J17544-2619, or IGR~J11215-5952
\citep[e.g.][]{walter06,sguera06,sidoli}. The new SFXT presented here, \0840, was 
discovered in the Vela region on May 15, 2006 with \I\ during a bright outburst
lasting about 15 minutes \citep{disc}. The candidate optical
counterpart was first tentatively identified as the supergiant,
Ob5Ib(f), HD 74194 star \citep{disc,masetti06} located at 3 kpc in the
Vela region \citep[e.g.][]{walborn,humphreys,schroeder}. Successively, using the
\S\ X-ray Telescope (XRT) a refined source position ($\alpha_{J2000} =
08^{\rm h}49^{\rm m}47^{\rm s}\!.97$ and 
$\delta_{J2000} = -45^{\circ}03^{\prime}29^{\prime\prime}\!\!.8$ with an uncertainty of $5^{\prime\prime}\!\!.4$) was derived, strengthening the association with HD 74194 \citep{swiftatel}.

Optical spectra of HD 74194 measured a few days after the outburst \citep{barba} strongly resemble the ones of the companion star of another SFXT, IGR J17544--2619 \citep{pellizza}.
In addition the radial velocity variations in the HeI and HeII absorption lines with an amplitude of
about 35 km s$^{-1}$ were measured, suggesting either a pulsating variable nature of the 
supergiant star HD 74194 or a possible Doppler orbital modulation \citep[e.g.][]{conti}. 
 
\0840 shows the typical SFXTs recurrent short flaring
events. Using \I\ archival data, an outburst was detected on July 1,
2003 \citep{mereghetti}, while 
the last bright flare was observed on October 4, 2006 with the \S~
satellite \citep{swiftgcn}.

In this letter we present the light curve and broad-band spectral study
of all the three outbursts of \0840\ observed up to now. In addition, thanks to the knowledge of the
companion star parameters, the possible wind accretion conditions are discussed.

\section{Data Analysis and Results}

\0840 was observed twice in outburst with \I\ \citep{integ}, on  July
1, 2003 and May 15, 2006, i.e. during the satellite
orbits 83 and 438. This dataset includes publicly available
data and part of AO3 Vela region guest observations. Hereafter we call the
two datasets flare 1 and flare 2, respectively. We used the data from the 
coded mask imaging telescope IBIS/ISGRI \citep{ibis,isgri} at energies
between 15 keV and 200 keV and from the JEM-X monitor \citep{jemx} between 3 and 20 keV. 
For JEM-X the data were extracted only for flare 2, flare 1 being
outside the JEM-X field of view. The data reduction was performed
using the standard Offline Science Analysis (OSA) version 5.1. 

Single pointings ($\sim 2000$ s each) were deconvolved and analyzed
separately.
The IBIS/ISGRI 15--40 keV high energy light curve,  shown in Fig. \ref{fig:b1lc},
has been extracted from the images using all available pointings. Light curves with a 100 s time
bin were also extracted around the outburst peaks.

The third outburst, flare 3, was observed on October 4, 2006
(14:45:42 UT) with \S/BAT \citep{swift,bat}. Since the flare was detected with an image trigger (15--150 keV), 
only ``Survey'' data products are available from the BAT instrument, with
a typical integration time of $\sim300$ s. We note that the BAT data
coverage is not continuous due to the low Earth orbit of the
\S\ satellite.
For each $\sim300$ s time bin the target count rate in the 14--40
keV energy range was evaluated with the mask-weighting technique using the XRT
coordinates. The BAT data were
selected 15 h before and 30 h after the trigger, and
reduced using the standard software version 2.5.

\S~ slewed automatically to the direction of the source and XRT \citep{xrt} (0.1--10 keV) follow-up observations were performed.
The XRT data were reduced using {\sc ftools} version 6.1.1., and processed with {\sc xrtpipeline} version 0.10.4, using
standard filtering selections. 
The XRT 0.1--10 keV energy band light curves have been extracted by
selecting a circular region of 20 pixels around the source for the
Photon Counting (PC) mode, and rectangular region of 40$\times$20
pixels for the Window Timing (WT) mode. Background light curves were
extracted in source free regions in order to produce the final
background-subtracted XRT/PC/WT light curves. The count rate of the source in
both modes is low enough in order to avoid the pile-up in the CCD.

\subsection{Flare light curves}

\begin{figure*}[ht!]
\centering
\includegraphics[height=18cm,width=7.5cm,angle=-90]{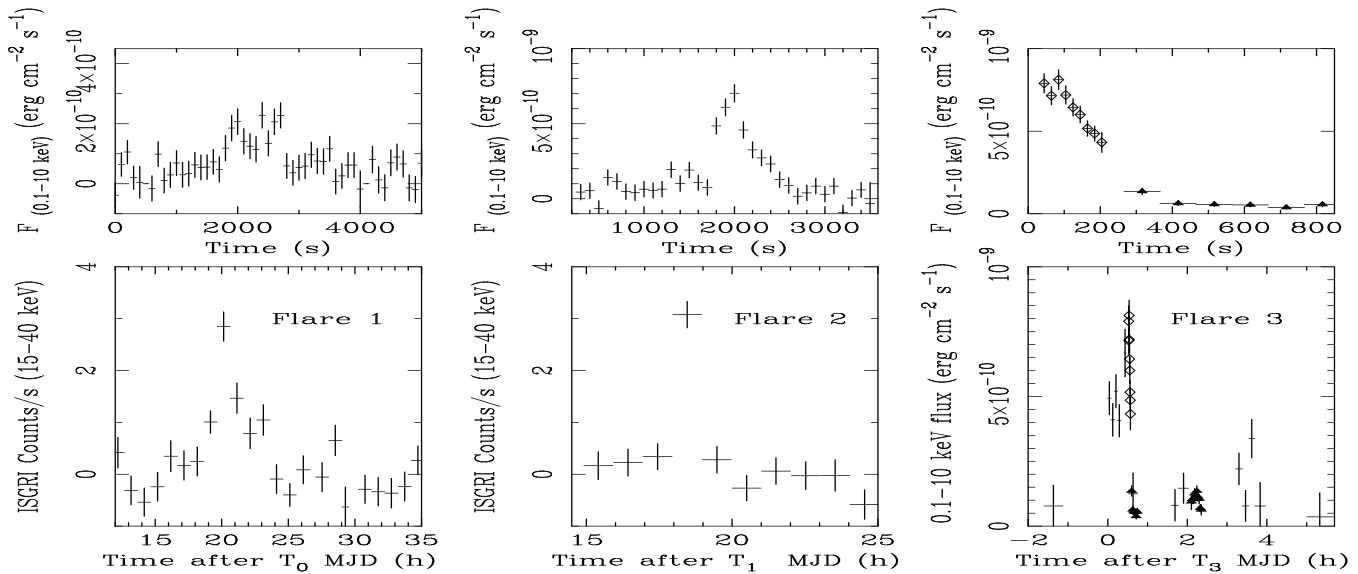}
\caption{{\em Bottom, from left:} IBIS/ISGRI (flare 1,2) and \S\ (flare 3) light curves. IBIS data are binned at $\sim 2$ ks. In the last panel the crosses represent the extrapolated BAT flux, while
the diamonds and  triangles the XRT/WT and PC data. The time axis is expressed in hours after T$_{0}$=52821.0, T$_{1}$=53870.0, and  T$_{3}$=54012.615 MJD, respectively. The \S\ light curve shows also weak flares 
e.g. at t$\sim$3.5 h \0840 is detected in BAT at $\sim$7 $\sigma$ level. 
{\em Top:} the brightest part of each outburst is shown on a smaller time scale (100 s time bin for ISGRI).}
\label{fig:b1lc}
\end{figure*}

In Fig. \ref{fig:b1lc} we show
the three outburst light curves of \0840 observed up to now. 
The light curves are dominated by a first bright flare that typically
reaches a peak luminosity of $3-6\times10^{36}$ erg
s$^{-1}$ (see Sec. \ref{sec:spec}) in a time scale of several tens of minutes to hours. The observed flares
vary from an isolated strong flare (flare 2) to a more
structured flaring activity (flare 1 and 3). Moreover flare 3 shows also
weaker flares with a luminosity of the order of $5\times10^{35}$
erg s$^{-1}$. 
The three outburst durations are of the
order of several hours.  The flare characteristics are similar to
those observed in other SFXTs, such as XTE~J1739-302
\citep[e.g.][]{sguera}, IGR~J16465-4507 \citep[e.g.][]{lutovinov05},
IGR~J17544-2691 \citep[e.g.][]{zand}, and IGR~J11215-5952
\citep[e.g.][]{sidoli}, but resemble also those well
studied in persistent wind accretion high mass X-ray binary (HMXB) sources like Vela~X-1
\citep[e.g.][]{white}, 4U~1700-377 \citep[e.g.][]{rubin96}. 

Up to now only IGR~J11215-5952 has shown 
a $\sim 330$ days periodic outburst. Therefore,
using a linear orbital function $T(n) = T_{0}+ nP_{\rm orb}$ we
 attempted to test the hypothesis of a periodically recurrent
outburst also for \0840. $P_{\rm orb}$ is the orbital period in days and
$T_{0}$ is taken from flare 2 (MJD 53780.771215). A possible
orbital period is 4.2988 days (with $n_{1}=244$ and
$n_{2}=33$), however, 
searching in the \I\ archival data around the expected outburst time, we did
not find any significant flux excess at the \0840 source position. We
conclude that the three observed outbursts are sporadic episodes of
accretion of matter fed from the wind of the supergiant companion star.

In order to directly compare the light curves measured with
different instruments (see Fig. \ref{fig:b1lc}), all the count rates have been converted to the
0.1--10 keV fluxes assuming the spectral shape derived in Sec. \ref{sec:spec}.

\subsection{Flare spectra}
\label{sec:spec}

We performed the spectral analysis using XSPEC version 11.3, for the ISGRI data (20--200 keV), flare 1,
for the 3.5--20 keV JEM-X data with the simultaneous 20--200 keV
ISGRI data, flare 2, and for the 0.3--100 keV simultaneous XRT/BAT
data, flare 3. Only the brightest parts of the flares have been considered.
For the simultaneous data a constant factor was
included in the fit to take into account the uncertainty in the
cross-calibration of the instruments.  
All data were rebinned in order to have
$3\sigma$ points and the spectral uncertainties in the results are given
at a 90\% confidence level for a single parameter. We use the source
distance of 3 kpc throughout the paper.

The XRT/BAT (flare 3) broad-band (0.3--100 keV) dataset provides
us the widest energy coverage and the best statistics, so we fitted it first.
We fitted the data using a simple photoelectrically-absorbed
power-law ({\sc pl}) model which was found inadequate with a
{\chiq}/d.o.f. = 132.76/76. 
The addition of a high-energy exponential cut-off significantly
improved the fit to {\chiq}/d.o.f.=80.36/75, resulting in a best-fit photon
index of 0.1$\pm$0.2 and a cut-off energy at 15$\pm$5 keV.  
The hydrogen column density, $N_{\rm H}$, was found to be $1.0\pm0.3\times10^{21}$
cm$^{-2}$. This value is compatible with
Galactic value in the direction of the source, $N_{\rm H}=3\times10^{21}$ cm$^{-2}$, reported in
the radio maps of \citet{nh}.   
We attempted to fit the low energy spectrum also with a thermal black-body
model, {\sc bb}, plus a {\pl} for the high energies, but we found high {\sc bb} temperature,
$kT_{\rm BB}\sim 8$ keV, which is much larger than the measured values for a neutron star (NS) surface or polar cap thermal emission.
Also a fit with a thermal bremsstrahlung model can
be statistically ruled out by our data ({\chiq}/d.o.f.=183.2/76). 
The energy spectra during the flaring activity are hence best fitted with a high-energy
exponential cut-off {\sc pl} model similar to the spectra observed from persistent wind accreting HMXB hosting a NS, e.g. Vela~X-1 \citep{white}. 

The joint JEM-X/ISGRI (3.5--200 keV), flare 2, 
spectrum  was also fitted with the cut-off {\sc pl} model used
for the flare 3. The best fit values are similar to the ones of flare 3, see Table \ref{tab:spectra}.
Given that we were not able to constrain the {\nh} value (as the JEM-X
bandpass starts above 3 keV) we fixed it to the value found from flare 3.

For flare 1 only ISGRI data (20--200 keV) are available.
The data can be well fitted with a simple {\sc pl} model with a photon index,
$\Gamma$, of 2.5$\pm$0.5. Note that fitting flare 2 (ISGRI) and 3 (BAT) spectra in the same energy range gives consistent photon index values, namely $\Gamma$=2.8$\pm$0.3 and $\Gamma$=2.3$\pm$0.4, respectively.
The apparently soft spectra derived using only the data above 20 keV show how important it is, to have a broad energy coverage in order to better characterize the spectrum of these sources.
To be coherent, we applied the same spectral model derived from the
broad-band fit of flare 2 and 3, also for flare 1, even if we had a smaller energy range.
We fixed the slope of the power law index to the values found (see Table 
\ref{tab:spectra}) and found the cut-off energy compatible with
the one found for flares 2 and 3.

In Fig. \ref{fig:b1lc}, last panel, weak flares are visible after the main peak. To infer the flux at
very low mass accretion rate we extracted a XRT/PC
spectrum in the 0.1--6 keV energy band at $\sim$2.2 hr after the main flare, see Fig. \ref{fig:b1lc}. This is the last time interval where the source is clearly detected in the XRT images.
We call this weak flare, flare 4. Using the same model found 
for flare 3, the source was at a bolometric luminosity of  $5\times10^{35}$
erg s$^{-1}$, with a marginal indication of an
increase of the absorption column density.

\begin{figure}[ht]
\centering
\includegraphics[width=5.0cm,angle=-90]{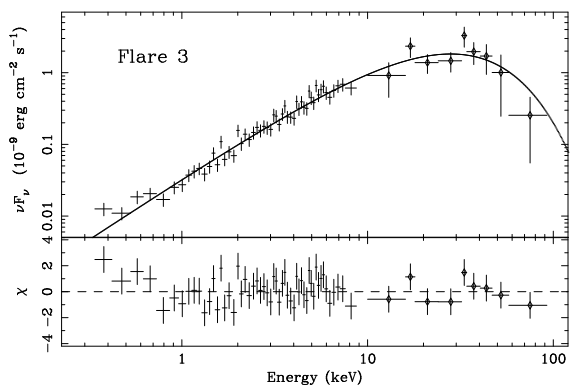}
\caption{Combined XRT (crosses) and BAT (diamonds) spectra of flare
  3. The best fit {\sc pl} high energy
cut-off model and the residuals (lower panel) are shown.} 
\label{fig:burst3}
\end{figure}

The best fit parameters for the different flares are reported in Table
\ref{tab:spectra}.  
In Fig. \ref{fig:burst3}, we show the broad-band unfolded
spectrum and the residuals of the data to the {\sc pl} with a high energy
cut-off model for flare 3.

\begin{deluxetable}{ccccc}
\tablecolumns{5}
\tablewidth{0pc}
\tablecaption{\0840 spectral parameters using a power-law model with an exponential
  high energy cut-off\label{fits}}
\tablehead{\colhead{}&\colhead{Flare 1} & \colhead{Flare 2} &\colhead{Flare 3} &\colhead{Flare 4}\\
\colhead{Dataset} & \colhead{ISGRI} & \colhead{JEM-X/ISGRI} &\colhead{XRT/BAT} &\colhead{XRT/BAT}}
\startdata
N$_{\rm H}$ \tablenotemark{a}& \nodata & 0.1 (f) & 0.1$\pm$0.03 & $<$0.75\\ 	
$\Gamma$ & 0.0 (f) & 0.0$\pm$0.5 & 0.1$\pm$0.2 &0.1$\pm$0.4\\
$E_{\rm c}$ (keV) & 12.7$\pm$2.8 & 11.8$\pm$2.8 & 15$\pm$5 & 15 (f)\\
$\chi^{2}$/d.o.f & 14.06/12 & 57.09/90 & 80.36/75 & 11.72/10\\
Flux\tablenotemark{b}$_{\rm bol}$ & 7.0$\times$10$^{-10}$ &
 2.7$\times$10$^{-9}$ & 6.0$\times$10$^{-9}$ & 5.0$\times$10$^{-10}$\\
\enddata
\tablenotetext{a}{in units of 10$^{22}$ cm$^{-2}$.}
\tablenotetext{b}{Unabsorbed 0.1--100 keV flux in units of erg cm$^{-2}$ s$^{-1}$.}
\label{tab:spectra}
\end{deluxetable}

\section{Discussion and Conclusions}
\label{sec:discussion}

\0840\ is a recurrent SFXT with most likely sporadic episodes of
accretion of matter from the wind of the  Ob5Ib(f), HD 74194, supergiant
companion star. The  broad-band spectrum (0.3--200 keV)
allowed us to perform an improved spectral analysis for the new source
\0840\ using XRT/BAT and JEM-X/ISGRI data. The best fit to the
data required a {\sc pl} with a high energy cut-off, see Table
\ref{tab:spectra}. The spectrum is typical for HMXBs hosting a NS
\citep{white} where the emergent radiation is presumably produced by
accretion columns at the magnetic poles of the NS. The bolometric
flare peak luminosity is of the order 
of $3-6\times10^{36}$ erg s$^{-1}$, which are typical values for wind accreting
HMXB. At low mass accretion rates, the source is observed at a bolometric
luminosity of $5\times10^{35}$ erg s$^{-1}$. The source quiescent
luminosity is in the order of $\sim 2\times10^{32}$ erg s$^{-1}$
\citep{swiftatel}. 

During the outbursts the derived absorption of $\sim 1\times10^{21}$
cm$^{-2}$ is compatible with the total Galactic absorption in that direction as
estimated from the HI maps, indicating that the source is not
surrounded by large amounts of material.  Some information about the
distribution of matter around an X-ray binary can be inferred from
observations of the variations with orbital phase of its spectrum
caused by absorption along the line of sight to the X-ray star. For
\0840, the orbital period is not known, therefore using a simply
spherically symmetric stellar wind model, we
approximate the wind density \citep[e.g. see Eq. (7) in][]{heap} at the NS surface ($a>>R_{*}$, $a$ is the
binary separation) for a circular orbit:
\begin{eqnarray}
 \lefteqn{N_{\rm H} \approx 1\times10^{21} \times}  \\
&&\biggl(\frac{\dot M_{w}}{10^{-7} M_{\odot} {\rm yr}^{-1}}\biggl)\biggl(\frac{R_{*}}{14.5R_{\odot}}\biggl)^{-1}\biggl(\frac{\nu_{\infty}}{2000\, {\rm km\,\,  s}^{-1}}\biggl)^{-1} {\rm cm}^{-2}, \nonumber
\label{nh}
\end{eqnarray}

where $\dot M_{w}$ is the HD 74194 mass
loss rate and $\nu_{\infty}$ is the terminal wind velocity. 
The used stellar parameters for HD 74194 are reported in Table
\ref{tab:star}. We found the estimated $N_{\rm H}$ compatible with the
derived absorption from the spectral fit. On the other hand, 
locating the companion star at $a=2R_{*}$, and using a maximal wind acceleration parameter $\beta=1$,
$N_{\rm H}$ will change by a factor ln($R_{*})\approx27$, i.e. $N_{\rm
  H}\approx2.7\times10^{22}$ cm$^{-2}$. The observed $N_{\rm H}$ hints
towards a large binary separation and therefore a large orbital period. 
The measured absorption is in variance with what has been observed for other SFXT \citep[see e.g.][]{walter06}. 

\begin{deluxetable}{cc}
\tablecolumns{2}
\tablewidth{0pc}
\tablecaption{Stellar parameters for HD 74194 derived from \citep{lamers}}
\tablehead{\colhead{Parameter} &\colhead{Value}}
\startdata
$T_{\rm eff}$ (k) & 33000\\
$R_{*}$ ($R_{\odot}$) & 14.5\\
$M_{*}$ ($M_{\odot}$) & 28\\
log($L_{*}/L_{\odot}$) & 5.3 \\
$\dot M_{w}$\tablenotemark{a} ($M_{\odot}$ yr$^{-1}$) & 2.3$\times10^{-7}$\\
$\nu_{\infty}$ (km s$^{-1}$)& 2000$\pm300$\\
$\nu_{\rm esc}$ (km s$^{-1}$) & 776\\
\enddata
\tablenotetext{a}{Calculated using Eq. \ref{mdot}.}
\label{tab:star}
\end{deluxetable}

The companion mass loss rate, $\dot M_{w}$, was calculated for  HD
74194 using the average relation for the galactic O or B supergiants
stars \citep{lamersbook},
\be
{\rm log}(\dot M_{w}) = -1.37 + 2.07\,{\rm log}(L_{*}/10^{6})-{\rm
  log}(\nu_{\infty}R_{*}^{0.5}).
\label{mdot}
\ee
The $\dot M_{w}$ value is a factor 10--100 lower that what found for
the persistent wind accretion HMXB, like Vela X-1 or 4U 1700-37. The
free parameter to obtain a higher $N_{\rm H}$ value for \0840\ is to have a higher mass loss
rates, $\dot M_{w}$. Using Eqs. (\ref{nh}) and (\ref{mdot}) we can see
that if the terminal wind velocity is a factor 5 lower (as observed
for an intrinsically obscured high energy source IGR J16318-4848
\citep{filliatre}), we would have a higher
absorption of $\sim10^{23}$ cm$^{-2}$, but such a low
wind velocity is not observed for HD 74194. This indicates that during the
accretion of matter for \0840\ there is no accretion wake of dense
matter surrounding the compact object \citep{blondin}.

We will assume that all the material within the capture radius 
\begin{eqnarray}
R_{\rm acc} & = &\frac{2 G M_{\rm x}}{\nu_{\rm x}^{2}+\nu_{\infty}^{2}}  \\
& \approx & 9.2\times10^{8}\biggl(\frac{\dot M_{acc}}{\dot
  M_{w}}\biggl)^{1/2}\biggl(\frac{M_{*}+M_{x}}{28M_{\odot}}\biggl)^{1/3}
P_{\rm orb}^{1/3} \,\,{\rm cm}\nonumber 
\label{orb}
\end{eqnarray}

is accreted by the compact object \citep{bondi}. $M_{\rm x}$ is the
mass of the NS, 
$\nu_{\rm x}$ its orbital velocity, and $\dot M_{acc}=
L_{x}c^{-2}\eta^{-1}$ the mass accretion rate, where $\eta\sim0.2$ is
the accretion efficiency for a NS. The rate of mass captured is 
then given by $\dot M_{\rm acc}=\dot M_{w}R_{\rm acc}^{2}/(4a^{2})$, 
with $R_{\rm acc} = 9.3\times10^{9}$ cm. We consider that $\nu_{\infty}
>> \nu_{\rm x}$ and $a>>R_{*}$. From the measured persistent emission L$_{x}\sim 2\times 10^{32}$ erg s$^{-1}$, we derive from Eq. (\ref{orb}) an orbital period of $\sim$1.5 years. 

The condition for accretion to take place is that the NS
magnetosphere radius is within the mass capture radius and the
corotation radius, i.e. $R_{Mag} \leq R_{\rm acc}$ and $R_{Mag} \leq
R_{\rm cor}$  \citep{illarinov}.
If we set  $R_{\rm acc}
= R_{\rm cor}$ we have $P_{\rm spin} =
2.36\times10^{27}(1.4M_{x})\nu_{\infty}^{-3} \simeq 7000$ s. The
magnetosphere radius is given by 
\begin{equation}
R_{Mag}=0.1\mu^{1/3}\dot M_{w}^{1/6}\nu_{\infty}^{-1/6}M_{*}^{1/9}P_{\rm orb}^{2/9}, 
\end{equation}
where $\mu = B R_{\rm NS}^{3}$.
Setting the accretion condition  $R_{Mag}=R_{\rm cor}=R_{\rm acc}$ the magnetic field has to be $B\sim 1.1\times 10^{13} (P_{\rm orb}/1 {\rm yr})^{-2/3}$ G, for $R_{\rm NS}=10^{6}$ cm. Assuming the orbital period derived above,
the NS magnetic field has to be of the order of 10$^{13}$ G. These are typical magnetic field values for young HMXBs hosting a NS, like e.g. Vela X-1 \citep{labarbera}.


The low $N_{\rm H}$ value measured during the flares is not consistent with the picture in which they are caused by clumps in the donor wind. In an alternative scenario, the flares could be associated with the sudden accretion onto the magnetic poles of matter previously stored in the magnetosphere during the quiescent phase. 
However, in order to have such a mass storage, the above simplest accretion conditions have to be studied for different scenarios (e.g. $R_{\rm acc} > R_{\rm cor} > R_{\rm mag}$). One can derive these conditions by varying opportunely $P_{\rm orb}$, $P_{\rm spin}$, and $B$ \citep{bozzo}. 

We conclude that these recurrent sporadic very short outburst episodes, due to the
accretion of matter from the wind of a supergiant companion star, imply
a spin period of the order of hours with a long orbital period, and a 10$^{13}$ G magnetic field for the NS.
The determination of all the system parameters can help to solve the accretion
mechanism.

\begin{acknowledgements}
\small{DG and MF acknowledge the French Space Agency (CNES) for financial support.
MF is grateful to Luigi Stella and Enrico Bozzo for helpful
discussions during his visit in Rome. ADL acknowledges an ASI fellowship. 
Based on observations with INTEGRAL, an ESA project with instruments and science data centre funded by ESA member states (especially the PI countries: Denmark, France, Germany, Italy, Switzerland, Spain), Czech Republic and Poland, and with the participation of Russia and the USA. We acknowledge the use of public data from the Swift data archive.}
\end{acknowledgements}


\begin{thebibliography}{}

\bibitem[\protect\citeauthoryear{Barba et al.}{2006}]{barba}
Barba, R., Gamen, R., \& Morrell, N. 2006,  Astr. Tel., 819

\bibitem[\protect\citeauthoryear{Barthelmy et al.}{2005}]{bat}
Barthelmy, S. D., et al. 2005, Space Science Reviews, 120, 143

\bibitem[\protect\citeauthoryear{Blondin}{1994}]{blondin}
Blondin, J., M. 1994, ApJ, 436, 756

\bibitem[\protect\citeauthoryear{Bozzo et al.}{2007}]{bozzo}
Bozzo, E., et al. 2007, in preparation

\bibitem[\protect\citeauthoryear{Burrows et al.}{2005}]{xrt}
Burrows, D.N., Hill, J.E, Nousek, J.A., et al. 2005, Space Sci. Rev., 120, 165

\bibitem[\protect\citeauthoryear{Bondi \& Hoyle}{1944}]{bondi}
Bondi, H., \& Hoyle, F. 1944, MNRAS, 104, 273

\bibitem[\protect\citeauthoryear{Conti et al.}{1977}]{conti}
Conti, P., S., Myckky Leep, E., Lorre, J., J. 1977, ApJ, 214, 759

\bibitem[\protect\citeauthoryear{Dickey \& Lockmann}{1990}]{nh}
Dickey \& Lockman, 1990, ARAA. 28, 215

\bibitem[\protect\citeauthoryear{Filliatre \& Chaty}{2004}]{filliatre}
Filliatre, P. \& Chaty, S., 2004, ApJ, 616, 469

\bibitem[\protect\citeauthoryear{Gehrels et al.}{2004}]{swift}
Gehrels, N., Chincarini, G., Giommi, P., et al. 2004, ApJ, 611, 1005

\bibitem[\protect\citeauthoryear{G\"otz et al.}{2006}]{disc}
G\"otz, D., Schanne, S., Rodriguez, J., et al. 2006,  Astr. Tel., 813

\bibitem[\protect\citeauthoryear{Heap \& Corcoran}{1992}]{heap}
Heap, S., R. \& Corcoran, M., F. 1992, ApJ, 387, 340

\bibitem[\protect\citeauthoryear{Humphreys}{1978}]{humphreys}
Humphreys, R., M., 1978, AJ, 38, 309

\bibitem[\protect\citeauthoryear{Illarinov, A. \& Sunyaev,}{1974}]{illarinov}
Illarinov, A. \& Sunyaev, R. 1977, A\&A, 39, 181

\bibitem[\protect\citeauthoryear{in't Zand}{2005}]{zand}
in't Zand, J.J.M. 2005, A\&A, 441, L1

\bibitem[\protect\citeauthoryear{Kennea \& Campana}{2006}]{swiftatel}
Kennea, J.A., Campana, S. 2006,  Astr. Tel., 818

\bibitem[\protect\citeauthoryear{La Barbera et al.}{2003}]{labarbera}
La Barbera, A., et al. 2003,  A\&A, 400, 993

\bibitem[\protect\citeauthoryear{Lamers et al.}{1995}]{lamers}
Lamers, H., J., G., L., M., Snow, T., P., \& Lindholm, D., M., 1995,
ApJ, 455, 269 

\bibitem[\protect\citeauthoryear{Lamers \& Cassinelli}{1999}]{lamersbook}
Lamers, H., J., G., L., M. \& Cassinelli, J., P. 1999, in Introduction
to Stellar Winds (Camb. Univ. Press) 

\bibitem[\protect\citeauthoryear{Lebrun et al.}{2003}]{isgri}
Lebrun,  F., Leray, J.-P., Lavocate, Ph., et al. 2003, A\&A, 411, L141

\bibitem[\protect\citeauthoryear{Lund et al.}{2003}]{jemx}
Lund, N., Budtz-Joergensen, C.,  Westgaard, N. J., et al.  2003, A\&A,
411, L231 

\bibitem[\protect\citeauthoryear{Lutovinov et al.}{2005}]{lutovinov05}
Lutovinov, A., Revnivtsev, M., Gilfanov, M., et al. 2005, A\&A, 444, 821

\bibitem[\protect\citeauthoryear{Masetti et al.}{2006}]{masetti06}
Masetti, N., et al. 2006,  Astr. Tel., 815

\bibitem[\protect\citeauthoryear{Mereghetti et al.}{2006}]{mereghetti}
Mereghetti, S., Sidoli, L., Paizis, A., \& G\"{o}tz, D. 2006, Astr. Tel., 814


\bibitem[\protect\citeauthoryear{Negueruela et al.}{2006}]{negueruela}
Negueruela, I., et al., ApJ, 638, 982

\bibitem[\protect\citeauthoryear{Pellizza et al.}{2006}]{pellizza}
Pellizza, L.J., Chaty, S., \& Negueruela, I. 2006, A\&A, 455, 653

\bibitem[\protect\citeauthoryear{Rubin et al.}{1996}]{rubin96}
Rubin, B., C., Finger, M., H., Harmon, B., A., 1996, ApJ, 459, 259

\bibitem[\protect\citeauthoryear{Schr\"{o}der et al.}{2004}]{schroeder}
Schr\"{o}der, S.E., et al. 2004, A\&A, 428, 149

\bibitem[\protect\citeauthoryear{Sidoli et al.}{2006}]{sidoli}
Sidoli, L., Paizis, A., \& Mereghetti, S. 2006, A\&A, 450, L9

\bibitem[\protect\citeauthoryear{Sguera et al.}{2005}]{sguera}
Sguera, V., Barlow, E.J., Bird, A.J., et al. 2006, A\&A, 444, 221

\bibitem[\protect\citeauthoryear{Sguera et al.}{2006}]{sguera06}
Sguera, V., Bazzano, A., Bird, A.J., et al. 2006, ApJ, 646, 452

\bibitem[\protect\citeauthoryear{Smith et al.}{2006}]{smith}
Smith, D.M., Heindl, W.A., Markwardt, C.B., et al. 2006, ApJ, 638, 974

\bibitem[\protect\citeauthoryear{Ubertini et al.}{2003}]{ibis}
Ubertini, P., Lebrun, F., Di  Cocco, G., et al. 2003, A\&A, 411, L131

\bibitem[\protect\citeauthoryear{Walborn}{1973}]{walborn}
Walborn, N. R. 1978, AJ, 78, 1067

\bibitem[\protect\citeauthoryear{Walter et al.}{2006}]{walter06}
Walter, R., Zurita Heras, J., Bassani, L., et al. 2006, A\&A, 453, 133

\bibitem[\protect\citeauthoryear{White et al.}{1983}]{white}
White, N.E., Swank, J.H., \& Holt, S.S. 1983, ApJ, 270, 711

\bibitem[\protect\citeauthoryear{Winkler et al.}{2003}]{integ}
Winkler, C., Courvoisier, T.J.-L., Di Cocco G., et al. 2003, A\&A, 411, L1

\bibitem[\protect\citeauthoryear{Ziaeepour et al.}{2006}]{swiftgcn}
Ziaeepour, H., Burrows, D.N., Campana, S., et al. 2006, GCN Circ., 5687

\end{thebibliography}
\end{document}